\documentclass{article}

\usepackage[dblblindworkshop,final]{neurips_2026}

\workshoptitle{AI for Drug Discovery (AI4DD)}

\usepackage{xcolor}         %
\definecolor{improve}{RGB}{0,80,45}
\definecolor{decline}{RGB}{140,30,30}

\usepackage[utf8]{inputenc} %
\usepackage[T1]{fontenc}    %
\usepackage{hyperref}       %
\usepackage{url}            %
\usepackage{booktabs}       %
\usepackage{amsfonts}       %
\usepackage{nicefrac}       %
\usepackage{microtype}      %
\usepackage{xcolor}         %
\usepackage{graphicx}
\usepackage[font=footnotesize,labelfont=bf,skip=2pt]{caption}
\usepackage{enumitem}
\usepackage{titlesec}
\usepackage{setspace}

\usepackage{amsmath}
\usepackage{tikz}
\usetikzlibrary{bayesnet,fit,positioning,arrows}
\usepackage{standalone}   %

\usepackage{booktabs}
\usepackage{pifont}
\usepackage{array}
\usepackage{makecell}

\title{D2ML: Models for Direct-to-Biology Small Molecule Screening}

\author{%
AkshatKumar Nigam \enspace
Neri Amara \enspace
\textbf{Hillary Dequina} \enspace
Dyana Kenanova \enspace
\textbf{Seth Malmersjo}\\[3pt]
\textbf{Fiorella Ruggiu} \enspace
\textbf{Raymond V. Fucini} \enspace
\textbf{Jack Sadowsky} \enspace
\textbf{Stig K. Hansen} \enspace
\textbf{Imran S. Haque}\\[8pt]
Kimia Therapeutics, Berkeley, CA 94710\\
\texttt{\{anigam,$\cdots$,ihaque\}@kimiatx.com}
}

\begin{document}

\maketitle
\vspace{-12pt}
\setstretch{1.0}
\setlength{\parskip}{5.5pt}
\titlespacing*{\section}{0pt}{1.35ex plus .2ex minus .2ex}{0.8ex plus .1ex}
\titlespacing*{\subsection}{0pt}{1.1ex plus .2ex minus .2ex}{0.6ex plus .1ex}
\section{Introduction}

Molecular property prediction is a central problem in machine learning for drug discovery.
Property prediction models (e.g., for protein--ligand binding) are often highly
data-limited, and their performance degrades substantially outside the training distribution
\citep{wallach2018memorization,fooladi2025ood}. Chemical property landscapes compound this challenge: most molecules may be inactive for a given problem, while compounds
nearby in chemical space can nevertheless differ substantially in activity, a phenomenon
known as an ``activity cliff'' \citep{maggiora2006activity,stumpfe2012activity}.

Small-molecule discovery searches enormous chemical spaces under economic
constraints: custom synthesis costs approximately US\$1000 per compound,
and design cycles take around 10 weeks.
Conventional search therefore operates in two phases. \emph{Hit
identification} uses methods like high-throughput screening or DNA-encoded libraries, which amortize
pre-established libraries across screens of multiple targets, to broadly
sample chemical space and identify initial hits. In \emph{hit-to-lead} and \emph{lead optimization}, medicinal chemists iteratively
synthesize close analogs to explore local structure--activity relationships
while limiting experimental cost.
As a result, activity datasets often suffer from a ``streetlight effect''
in which active compounds occur in closely related series --- not because
they are the only possible actives, but because they are particularly accessible
through local search \citep{blevins2025clever}.
Models must therefore learn from data that are scarce, chemically biased, and
subject to temporal distribution shift during discovery program progress. Useful models must operate within a design--make--test--learn (DMTL) cycle in which experimental resources
are limited and compounds are evaluated against multiple objectives rather than a single endpoint.

An alternative strategy combines nanogram-scale
high-throughput chemical synthesis (HTCS) with biological testing of crude
reaction mixtures (i.e., taking reactions ``direct to biology'' without
purification), allowing ${>}10^{4}$ novel compounds to be made and tested in
days.
Carmot Therapeutics was the first to industrialize this HTCS direct-to-biology strategy (``D2B'' for short) as a centerpiece chemical exploration strategy, successfully discovering multiple drugs using the approach in a purely empirical experimentally-driven, rather than model-guided, manner \citep{hansen2018methods, shin2019discovery, rodriguez2025biased}.
Machine learning has also been applied to D2B
workflows to identify potent compounds obscured by low reaction
yields and to guide prospective selection from virtual libraries
\citep{mccorkindale2024deconvoluting}. Extending model-guided D2B
to an assay cascade introduces additional challenges. Even conventional chemical search requires modeling variable-fidelity data from high-throughput screening vs focused low-throughput assays; benefiting from D2B's scale further requires handling unpurified compounds with uncertain true concentrations. 

In this work, we investigate how molecular property prediction models can use
structured supervision from a D2B assay sequence. We tailor each prediction task to its assay readout and investigate
whether sharing information across stages improves prediction
when high-fidelity measurements are scarce.
We evaluate predictive performance on held-out assay data and
separately demonstrate model-guided compound discovery through a prospective
campaign. This campaign identifies compounds with improved potency and
selectivity in a large, synthetically accessible chemical space.

\section{Methods}

\begin{figure}[t]
  \centering
  \includegraphics[width=0.89\textwidth]{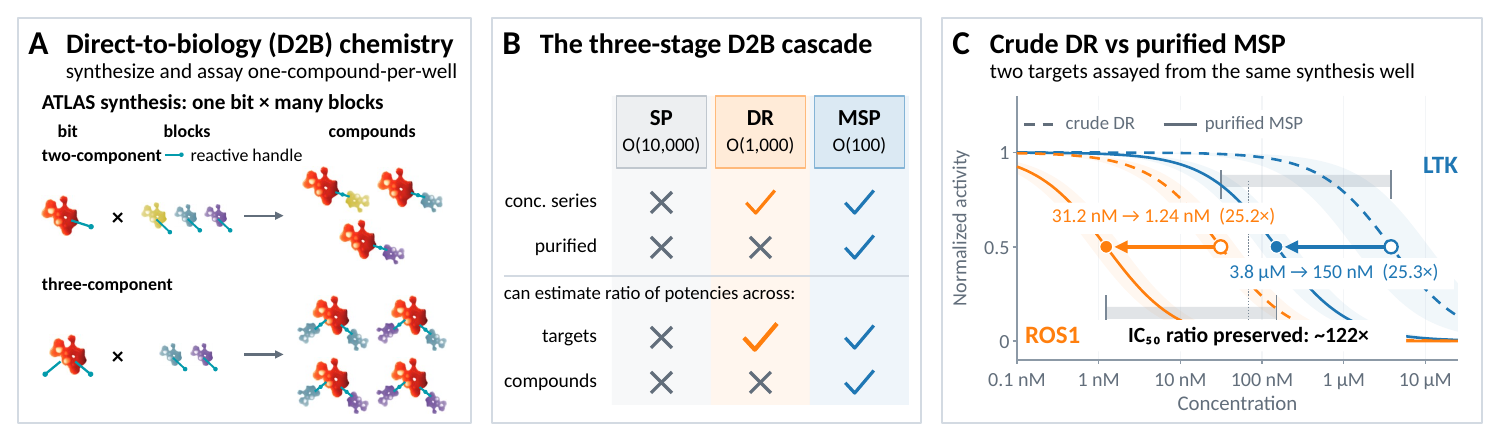}
  \caption{\textbf{Direct-to-biology synthesis, screening stages, and preservation of selectivity
  across purification.}
  \textbf{(A)} Kimia's ATLAS platform combines a target-directed \emph{bit} with libraries of
  \emph{blocks} in one-reaction-per-well format. Two-component reactions explore one molecular
  vector; three-component reactions explore two, yielding up to $n^2$ products from $n$ blocks.
  \textbf{(B)} The D2B cascade progresses from SP ($O(10^4)$ compounds) to DR ($O(10^3)$) and
  purified MSP ($O(10^2)$). DR and MSP support potency ratios across targets, while only MSP
  supports quantitative comparison across compounds.
  \textbf{(C)} Purification shifts ROS1 and LTK IC$_{50}$ values from the same crude well by
  nearly identical factors ($25.2\times$ and $25.3\times$), preserving the ${\sim}122$-fold
  potency ratio and showing that crude DR can retain selectivity even when absolute potency is
  uncertain. Supplementary Figure~\ref{fig:dr-msp-selectivity} shows a systematic paired analysis of DR/MSP potency ratios.} 
  \label{fig:cascade}
\end{figure}

\subsection{High-throughput chemical synthesis with direct-to-biology screening (D2B)}

Kimia's D2B synthesis and screening platform, ATLAS \citep{sadowsky2026d2b}, uses
standardized coupling reactions to link reactive core scaffolds (``bits'') with libraries of
complementary fragments (``blocks'') in an arrayed format (one-per-well;
Figure~\ref{fig:cascade}A) --- for example, linking carboxylic-acid-containing bits with
amine-containing blocks via amide coupling, or an aryl halide bit with a boronic acid block in a
Suzuki reaction. In a two-component reaction, a bit with a single reactive handle can be
cross-reacted against hundreds or thousands of blocks, one-per-well, to enable broad exploration
at a single molecular vector in one step. Three-component reactions leverage scaffolds with
multiple independently reactive handles to enable multi-vector exploration and search over
chemical spaces of $>10^8$ compounds by active learning in batches of $\mathrm{10^3-10^4}$ compounds.

ATLAS achieves its scale by combining miniaturized nanogram-scale chemical synthesis with
biological testing of \emph{unpurified} (crude) reaction mixtures; similar D2B approaches have
been used by Carmot and others in degrader, fragment, agonist, and inhibitor discovery campaigns
across covalent and noncovalent modalities
\citep{stevens2023integrated,wilders2025expedited,stevens2025expanding,
douthwaite2026cdk2,shin2019discovery,du2024usp28,rodriguez2025biased}. D2B lends itself to the sequential assay cascade shown in Figure~\ref{fig:cascade}B:

\begin{itemize}[topsep=2pt,itemsep=1pt,parsep=0pt,leftmargin=1.2em]
  \item \textbf{Crude single-point (SP):} each compound is first tested at a single, high
  concentration. Many wells are expected to be inactive due to broad exploration or failed synthesis. The SP screen cheaply identifies putatively active wells and is best interpreted as coarse evidence of activity rather
  than quantitative potency.

  \item \textbf{Crude dose--response (DR):} compounds passing SP screening are tested at multiple
  dilution ratios (concentrations); true actives should display
  concentration-dependent activity. Because these products are unpurified, the target molecule's exact concentration is unknown. However, when multiple targets are assayed
  using material from the same crude well, this concentration uncertainty is shared between
  measurements and cancels when their apparent potencies are compared. Crude DR can therefore
  preserve \emph{selectivity} even when absolute potency is uncertain. Figure~\ref{fig:cascade}C
  illustrates this directly: purification shifts the LTK and ROS1 dose--response curves by
  approximately the same factor while preserving their roughly 122-fold potency ratio. This
  interpretation assumes that other components of the crude reaction mixture do not themselves
  contribute substantial target-specific activity \citep{mccorkindale2024deconvoluting}. Across structure-matched DR/MSP compounds, selectivity is positively correlated for all three target pairs, suggesting that crude DR measurements can provide useful estimates of selectivity for compounds advanced to MSP (Supplementary Section~\ref{si:paired-selectivity} and Figure~\ref{fig:dr-msp-selectivity}).

  \item \textbf{Microgram-scale synthesis with purification (MSP):} selected compounds from DR are resynthesized at the $\mu$g scale, versus the ng scale used for SP and DR,
  then purified and quantitated. Purified material is assayed in dose--response at known
  concentration, yielding MSP data. These measurements provide the
  highest-fidelity potency estimates for a given compound, but at substantially decreased
  throughput.
\end{itemize}

\begin{figure}[t]
  \centering
  \includegraphics[width=0.89\textwidth]{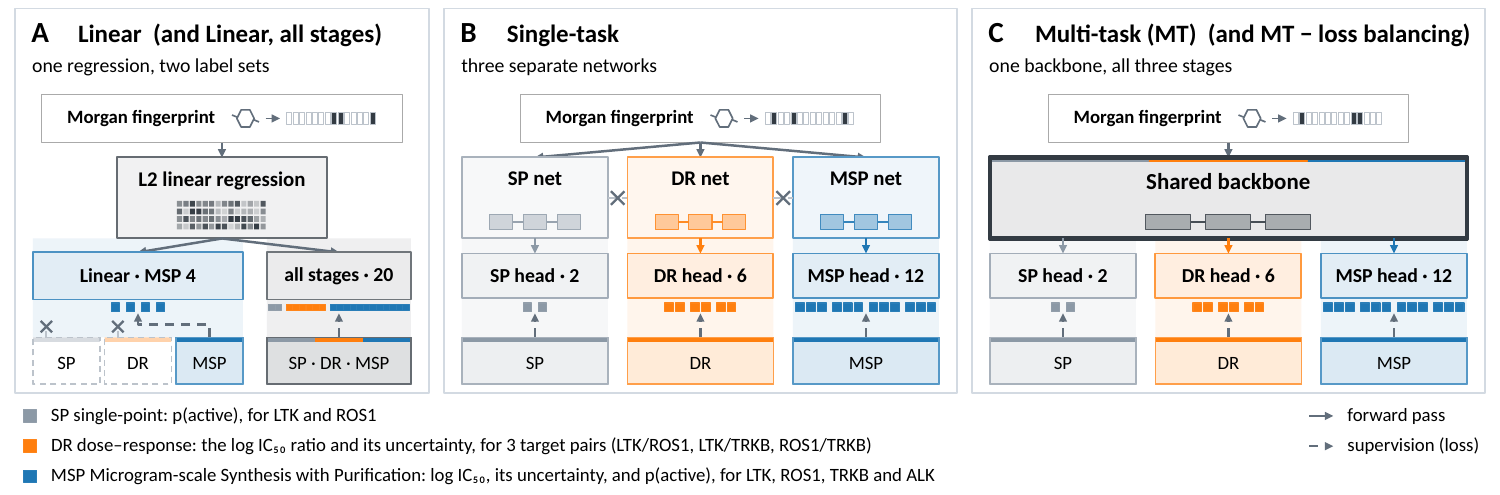}
  \caption{\textbf{Model architectures across the D2B assay cascade.} All models take Morgan fingerprints and differ in supervision and information sharing.
  \textbf{(A)} L2-regularized linear regression, one coefficient matrix: ``Linear'' is fitted to
  MSP alone, and ``Linear (all stages)'' to all three stages. \textbf{(B)} Single-task: independent networks for each stage. \textbf{(C)} Multi-task: a shared backbone with
  stage-specific heads; ``MT $-$ loss balancing'' removes stage-wise loss normalization and
  weighting. Solid arrows denote the forward pass; dashed arrows denote supervision.}
  \label{fig:arch}
\end{figure}

\subsection{Model design for D2B}

We ask whether SP/DR data should only guide selection for MSP or also
be used to train predictive models. Put
differently, does the structured noise arising from the lack of purification in SP/DR data
overwhelm the benefit of having 100--1000$\times$ the compounds going into a model? Chemical intuition suggests that measurements on impure compounds may be
uninformative or harmful, whereas larger datasets may overcome noise
\citep{halevy2009}.

We characterize a series of modeling strategies, illustrated in Figure~\ref{fig:arch}. Our focus is
on domain-specific model architecture rather than chemical representation, so we use Morgan
(ECFP-style) fingerprints as a common molecular representation across all models
\citep{rogers2010extended}. As a baseline, we evaluate L2-regularized linear regression (ridge
regression) in both a single-task setting purely on predicting MSP potency and a lumped setting
regressing the activity data for SP, DR, and MSP together, without separate losses, to test whether
access to the broader cascade is sufficient without a higher-capacity shared representation. We
additionally evaluated ordinary least squares (OLS) and L1 regularization (lasso), which yielded
qualitatively similar conclusions.

The main models of interest are single-task versus multi-task deep networks. In both settings, we
tailor custom loss heads for each assay stage: predicting activity probability at SP, pairwise log
potency ratios at DR, and absolute log IC$_{50}$ at MSP. Prior work has addressed yield--activity
confounding in D2B \citep{mccorkindale2024deconvoluting} and
transfer learning between screening fidelities
\citep{buterez2024transfer}. Our focus is on tailoring prediction
targets and losses to the information available at each D2B
stage. In particular, the crude DR objective uses cross-target
log potency ratios, in which the shared concentration factor
cancels, allowing selectivity to be modeled despite uncertainty
in absolute compound concentration. In the single-task setup, we train an
independent network for each stage; in the multi-task setup, we train the three stages jointly using
a shared molecular backbone and stage-specific output heads, with stage-wise loss normalization
and weighting to prevent differences in raw loss scale from dominating the shared objective
(Supplementary Section~\ref{si:balancing}). We compare this model with a separately tuned
multi-task configuration without loss balancing. Hyperparameters were optimized independently for each model over 40 Optuna trials
(Supplementary Section~\ref{si:hpo}).

\subsection{Dataset Design}

The data in this work come from a kinase discovery program targeting ROS1, a therapeutic target in
non-small-cell lung cancer. Libraries were constructed from a two-handle 2,4-disubstituted
pyrimidine bit using Kimia-proprietary regioselective three-component chemistry
(Figure~\ref{fig:cascade}A). Compounds were assayed for activity against four kinases: ROS1, its
close structural relatives ALK and LTK, and TRKB, which serves as a key anti-target. Selectivity
over TRKB is important because TRK-family inhibition can cause neurological toxicities
\citep{drilon2023nvl520}. The program therefore seeks compounds potent against ROS1 and selective over TRKB, making this a multi-objective design problem. Activity
was tested using ADP-Glo (Promega) at $25\,\mu\mathrm{M}$ ATP.

Table~\ref{tab:dataset} describes the composition of the dataset by assay stage. All models were
evaluated on retrospective test sets constructed to maximize separation from the training data
(Supplementary Section~\ref{si:splits}). For MSP, we report two complementary evaluations:
MSP-held is the resulting out-of-distribution retrospective set of 24 molecules; MSP-prospective is
a separate set of 84 molecules derived from a prospective experiment and characterized with
high-fidelity measurements.

\begin{table}[t]
\centering
\footnotesize
\setlength{\tabcolsep}{6pt}
\renewcommand{\arraystretch}{1.15}

\begin{tabular}{lrrrrcrrrr}
\toprule
& \multicolumn{4}{c}{Compounds assayed per target}
& & \multicolumn{4}{c}{Compounds per split} \\
\cmidrule(lr){2-5}\cmidrule(lr){7-10}
Stage & LTK & ROS1 & TRKB & ALK & & train & val. & test & total \\
\midrule
SP              & 26,538 & 15,368 & --    & --  & & 21,228 & 2,654 & 2,656 & 26,538 \\
DR              &  1,462 &  1,332 & 1,163 & --  & &  1,273 &   163 &   180 &  1,616 \\
MSP             &    163 &    163 &   163 & 163 & &    114 &    25 &    24 &    163 \\
\addlinespace[2pt]
MSP-prospective &     84 &     84 &    84 &  84 & &     -- &    -- &    84 &     84 \\
\bottomrule
\end{tabular}

\vspace{4pt}

\caption{\footnotesize Dataset used in this work. Left: per-target assay coverage; right:
train--validation--test counts. Each compound has one split shared across assay stages, preventing
cross-stage leakage (Supplementary Section~\ref{si:splits}). The 24-compound MSP test split is
MSP-held (Table~\ref{tab:generations}); MSP-prospective is a separate prospective set used only for evaluation.}
\label{tab:dataset}
\end{table}

\section{Results}

\begin{table}[t]
\centering
\setlength{\tabcolsep}{3.5pt}
\renewcommand{\arraystretch}{1.3}

\resizebox{\textwidth}{!}{%
\begin{tabular}{l
                r@{\,}c@{\,}l
                r@{\,}c@{\,}l
                r@{\,}c@{\,}l
                r@{\,}c@{\,}l
                r@{\,}c@{\,}l
                r@{\,}c@{\,}l
                r@{\,}c@{\,}l}
\toprule
& \multicolumn{3}{c}{SP}
& \multicolumn{6}{c}{DR}
& \multicolumn{6}{c}{MSP-held}
& \multicolumn{6}{c}{MSP-prospective} \\
\cmidrule(lr){2-4}
\cmidrule(lr){5-10}
\cmidrule(lr){11-16}
\cmidrule(lr){17-22}

Model
& \multicolumn{3}{c}{ROC-AUC}
& \multicolumn{3}{c}{$\rho$}
& \multicolumn{3}{c}{$r$}
& \multicolumn{3}{c}{$\rho$}
& \multicolumn{3}{c}{$r$}
& \multicolumn{3}{c}{$\rho$}
& \multicolumn{3}{c}{$r$} \\
\midrule

Linear
& \multicolumn{3}{c}{--}
& \multicolumn{3}{c}{--}
& \multicolumn{3}{c}{--}
& \multicolumn{3}{c}{$0.624$}
& \multicolumn{3}{c}{$0.616$}
& \multicolumn{3}{c}{$-0.029$}
& \multicolumn{3}{c}{$-0.021$} \\

\addlinespace[2pt]

Linear (all stages)
& \multicolumn{3}{c}{$0.732$}
& \multicolumn{3}{c}{$0.241$}
& \multicolumn{3}{c}{$0.174$}
& \multicolumn{3}{c}{$-0.011$}
& \multicolumn{3}{c}{$-0.117$}
& \multicolumn{3}{c}{$0.349$}
& \multicolumn{3}{c}{$0.357$} \\

\midrule

\addlinespace[2pt]

Single-task
& $0.890$ & $\pm$ & $.006$
& $0.271$ & $\pm$ & $.049$
& $0.259$ & $\pm$ & $.063$
& $0.420$ & $\pm$ & $.026$
& $0.365$ & $\pm$ & $.033$
& $0.120$ & $\pm$ & $.034$
& $0.150$ & $\pm$ & $.027$ \\

Multi-task (MT)
& $\mathbf{0.892}$ & $\pm$ & $.011$
& $\mathbf{0.402}$ & $\pm$ & $.024$
& $\mathbf{0.400}$ & $\pm$ & $.025$
& $\mathbf{0.443}$ & $\pm$ & $.047$
& $\mathbf{0.443}$ & $\pm$ & $.062$
& $\mathbf{0.175}$ & $\pm$ & $.015$
& $\mathbf{0.188}$ & $\pm$ & $.022$ \\

MT -loss balancing
& $0.874$ & $\pm$ & $.006$
& $0.336$ & $\pm$ & $.020$
& $0.334$ & $\pm$ & $.020$
& $0.427$ & $\pm$ & $.048$
& $0.357$ & $\pm$ & $.044$
& $0.133$ & $\pm$ & $.025$
& $0.132$ & $\pm$ & $.025$ \\

\bottomrule
\end{tabular}}

\vspace{4pt}

\caption{\textbf{Predictive performance across the D2B assay cascade.} SP is scored by macro
AUROC (LTK, ROS1); DR and MSP by Spearman ($\rho$) and Pearson ($r$), computed per target and
macro-averaged (three pairwise ratios at DR, four kinases at MSP). MSP-held is the retrospective
out-of-distribution set ($24$ molecules), MSP-prospective the independent prospective set ($84$).
The MSP-prospective compounds were selected using single-task models;
multi-task results on this cohort assess prediction accuracy, not the ability of multi-task
models to select compounds prospectively.
DR and MSP evaluation is restricted to resolved potency estimates
($p_{\mathrm{inact}} < 0.30$). Deep models report mean $\pm$ s.d. over five initializations;
linear baselines are single fits.}
  \label{tab:generations}
\end{table}

\subsection{Joint modeling improves prediction at sparse assay stages}

We asked whether information could be shared across the assay cascade to improve prediction at the
more sparsely sampled DR and MSP stages. The training sets span a near $200$-fold range in size,
from $21{,}228$ molecules at SP to $1{,}273$ at DR and only $114$ at MSP
(Table~\ref{tab:dataset}).

Table~\ref{tab:generations} reports predictive performance across the cascade for all five
models. Linear baselines demonstrate the difficulty of learning from the sparse MSP data:
although L2-regularized linear regression appears highly competitive on MSP-held ($\rho=0.624$, $r=0.616$),
it completely fails to generalize to MSP-prospective. All-stage ridge shows inconsistent generalization, performing poorly on MSP-held ($\rho=-0.011$, $r=-0.117$), but well on MSP-prospective ($\rho=0.349$, $r=0.357$).
Thus, while earlier-stage supervision \emph{may} benefit simple models, access to additional data alone does
not yield robust performance across assay stages and evaluation sets.

The deep model sequence provides a clearer benefit from specialization of losses and sharing representations. The single-task deep network generalizes
within each prediction modality (SP, DR, and MSP), with
performance falling off in a correlated manner to dataset size, suggesting that the MSP model in particular might benefit from
sharing parameters with earlier stages. Consistent with this hypothesis, a multi-task model dominates the single-task model on all endpoints; although the difference is small at SP (AUROC 0.892 vs 0.890), it grows materially at DR and MSP, reaching $\rho=0.443$ on MSP-held and $\rho=0.175$ on MSP-prospective.

Stage-specific losses can differ substantially in
scale, motivating the use of stage-wise loss normalization and
weighting (Supplementary Section~\ref{si:balancing}). An exploratory
comparison using BCE/Huber losses illustrates how balancing
redistributes the stage contributions to the training loss
(Supplementary Figure~\ref{fig:reconstructed-stage-losses}).
The multi-task configurations with and without loss balancing in
Table~\ref{tab:generations} were tuned independently and differ
in their molecular representations, network architectures, and
checkpoint-selection criteria; their performance differences
cannot be attributed to loss balancing alone.

\begin{figure}[t]
  \centering
  \includegraphics[width=0.89\textwidth]{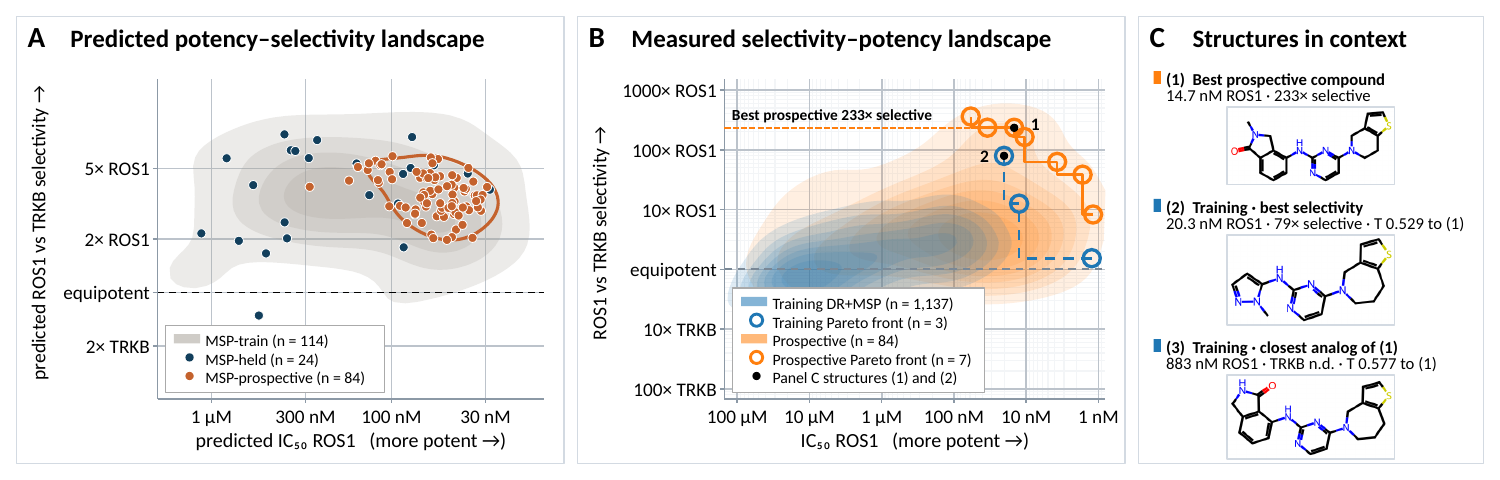 }
\caption{\footnotesize \textbf{Model-guided prospective selection occupies favorable MSP prediction space and extends the measured ROS1 potency--selectivity frontier.}
\textbf{(A,B)} Predicted (A) and measured (B) ROS1 potency and ROS1/TRKB selectivity ($\mathrm{IC_{50}}$ ratio). A: MSP-train prediction density (gray); MSP-held (blue) and -prospective (orange) predictions. B: DR/MSP training measurements (one per compound; $\mathrm{n=1{,}137}$, blue); MSP-prospective ($\mathrm{n=84}$, orange). Open circles denote Pareto-optimal compounds in either set; black markers identify compounds in panel C.
\textbf{(C)} The most selective prospective (1) and training (2) compounds
and the closest training analog of compound 1 (3). Changes are largest in the left-side substituent; (1) is \emph{not} a local change from (2), and its closest neighbor is weak, indicating exploration beyond local transformations of top hits (T = ECFP4 Tanimoto similarity).}
 
\label{fig:best}
\end{figure}

\subsection{Prospective validation extends the ROS1 potency--selectivity frontier}

\begin{figure}[t]
  \centering
  \includegraphics[width=0.50\textwidth]{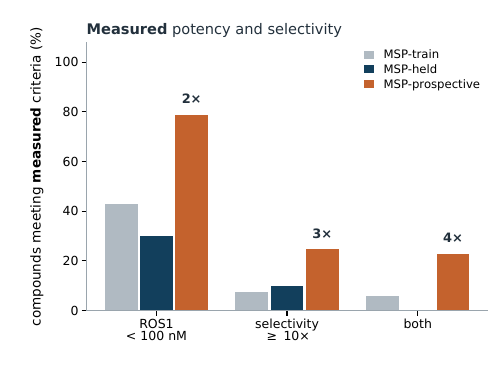}
\caption{\footnotesize \textbf{Prospective selection enriches measured ROS1 potency and
selectivity.}
Percentage of compounds satisfying measured criteria for ROS1 potency
(IC$_{50}<100$\,nM), ROS1-over-TRKB selectivity ($\geq10\times$), or both, in
MSP-train, MSP-held, and MSP-prospective. Only compounds with resolved potency estimates
for both targets ($p_{\mathrm{inact}}<0.30$) are included. Annotations report fold enrichment of MSP-prospective relative
to MSP-train. The corresponding distributions in
MSP prediction space are shown in Figure~\ref{fig:best}A.}
\label{fig:selection-vs-ranking}
\end{figure}

Before developing the multi-task model, we prospectively tested whether
single-task models could extend the observed potency--selectivity frontier. We enumerated commercially available Enamine blocks
reaction-compatible with our chosen 2,4-disubstituted pyrimidine bit to
generate a synthetically accessible virtual library of approximately $10^8$ molecules. We retained candidates predicted to (i) be active against ROS1 at SP, (ii) have at least
3-fold ROS1-over-TRKB selectivity at DR, (iii) retain at least 2-fold ROS1-over-TRKB selectivity
at MSP, and (iv) have a predicted MSP ROS1 IC$_{50}$ below 175\,nM. These filters guided fragment purchases to synthesize and test 84 novel compounds (MSP-prospective).
Table~\ref{tab:generations} evaluates prediction--measurement
agreement within this already selected set, whereas the combined
SP--DR--MSP selection pipeline targets enrichment in the favorable
tail. Weak within-set correlations can therefore
coexist with separation in MSP prediction space
(Figure~\ref{fig:best}A) and enrichment in measured potency and
selectivity (Supplementary Figure~\ref{fig:selection-vs-ranking}).

Prospective compounds extended the observed ROS1 potency--selectivity landscape
(Figure~\ref{fig:best}B). The best prospective compound (1) achieved 233-fold
ROS1-over-TRKB selectivity while retaining a ROS1 IC$_{50}$ of 14.7\,nM, compared with
79-fold selectivity and a ROS1 IC$_{50}$ of 20.3\,nM for the most selective compound in
training (Figure~\ref{fig:best}C). Compound 1 is only moderately similar to this training
compound (ECFP4 Tanimoto $0.529$), indicating that the improvement was not obtained through
local elaboration of the strongest existing chemotype. Its closest historical analog (3; Tanimoto $0.577$) was
substantially weaker against ROS1 (IC$_{50}=883$\,nM),
with unresolved TRKB potency. The prospective compounds extend the observed Pareto frontier toward
higher selectivity while retaining potent activity (Figure~\ref{fig:best}B).
Restricting the comparison to purified MSP measurements changes
historical density but preserves both Pareto frontiers, since all
historical frontier compounds were measured by MSP (Supplementary
Section~\ref{si:purified-frontier} and Figure~\ref{fig:pareto-msp-only}).
The historical MSP cohort was already enriched for activity through experimental
SP/DR screening, whereas the prospective compounds were selected using model
predictions and tested directly by MSP.

\section{Conclusions and Future Work}

Matching prediction targets to assay information and sharing supervision
across stages can improve learning from heterogeneous D2B measurements. In prospective validation, the single-task D2B model guided exploration
of a large accessible chemical space to extend the observed ROS1/TRKB
potency--selectivity frontier, and a multi-task model further improved
retrospective performance. These benchmarks support balanced joint modeling while highlighting the
need to match model capacity to data availability. Future work should evaluate multi-task models prospectively, investigate
additional representations, test additional experimental confounds like active
bits, and explore how stage-specific assay supervision can be applied to
chemical foundation models. See Supplementary Section~\ref{si:limitations} for limitations.

\clearpage
\setstretch{1}
\setlength{\parskip}{5.5pt}
\titlespacing*{\section}{0pt}{2.0ex plus .5ex minus .2ex}{1.5ex plus .3ex minus .2ex}
\titlespacing*{\subsection}{0pt}{1.8ex plus .5ex minus .2ex}{0.8ex plus .2ex}
{
\small
\bibliographystyle{plainnat}
\bibliography{references}
}

\newpage

\section{Supplementary material}

\subsection{Bayesian dose--response curve fitting}
\label{si:curves}

\begin{figure}[b]
\includegraphics[width=\linewidth]{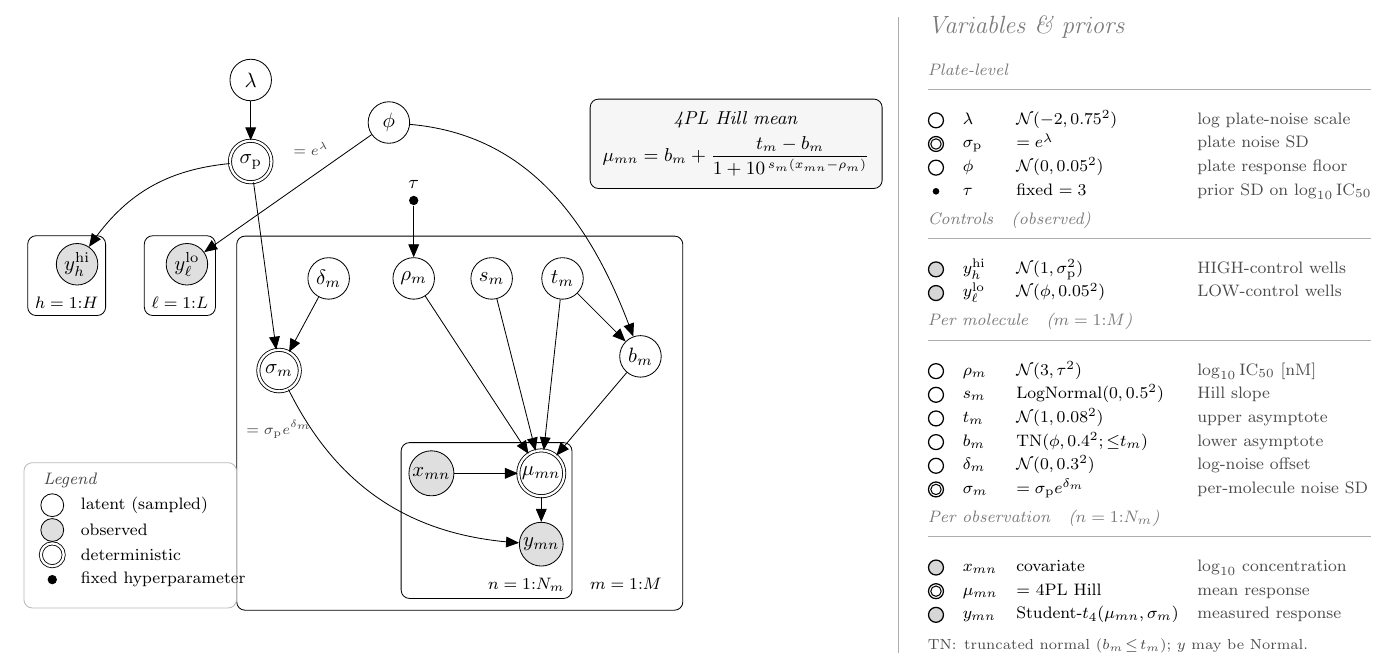}
\caption{Bayesian model for dose-response curve fitting. Robust Bayesian fits and parameter-sharing at the plate level avoid common numerical instabilities arising in least-squares fitting of assay data that may range from highly active to totally inactive, and enable downstream propagation of uncertainties. }
\label{fig:bayesdr}
\end{figure}

Dose--response measurements were analyzed using Bayesian curve fitting so that each measurement
was represented by both a potency estimate and its associated uncertainty, rather than by a single
least-squares point estimate. Raw well measurements were first normalized using plate-level controls,
with high and low controls defining the response scale. Curves were then fit jointly within each
plate using a four-parameter Hill model, with plate-level parameters capturing shared experimental
variation and molecule-specific parameters describing potency, slope, asymptotes, and residual
noise (Figure~\ref{fig:bayesdr}). A Student-$t$ likelihood was used to reduce sensitivity to outlier wells. Posterior
distributions were sampled using the No-U-Turn sampler \citep{hoffman2014no} implemented in
NumPyro \citep{phan2019composable}. For each dose--response curve, we retained the posterior median
of $\log_{10}\mathrm{IC}_{50}$ as the potency estimate and its posterior standard deviation as a
measure of uncertainty. When measurements were available across multiple plates or synthesis
batches, posterior samples were combined before calculating these summaries so that disagreement
between replicate measurements was reflected in the resulting uncertainty.

Because a numerical IC$_{50}$ can be returned even when the experimental data do not resolve a
dose--response relationship, we additionally estimated the probability that each compound was
inactive, $p_{\mathrm{inact}}$. This probability was derived from two properties of the fitted
posterior: the width of the credible interval on $\log_{10}\mathrm{IC}_{50}$, which captures how
well potency is identified, and the posterior curve depth, which captures evidence for a
non-flat dose--response relationship. These quantities were combined using a logistic classifier
calibrated against independent activity classifications. The classifier achieved a held-out
ROC-AUC of $0.933$, compared with $0.893$ when using credible-interval width alone. Measurements
with $p_{\mathrm{inact}} < 0.30$ were considered to have a resolved potency estimate and were used
for potency-based training and evaluation.

The resulting posterior summaries define the supervision used by the dose--response models. At
MSP, each target is represented by its absolute $\log_{10}\mathrm{IC}_{50}$, posterior uncertainty,
and activity probability ($1-p_{\mathrm{inact}}$). At DR, where uncertainty in the concentration
of crude material makes absolute potency unreliable, supervision is instead defined using pairwise
differences in $\log_{10}\mathrm{IC}_{50}$ between targets, corresponding to log potency ratios.
Uncertainty in each ratio is propagated from the posterior uncertainties of the two contributing
curves, and a ratio is used only when both curves have resolved potency estimates.

\subsection{Construction of the evaluation sets}
\label{si:splits}

To evaluate generalization to chemically distinct compounds, each compound was assigned a single
train--validation--test split that was shared across all three assay stages. The assignment was
constructed over the union of compounds appearing at SP, DR, or MSP ($27{,}161$ unique compounds)
and then applied to each stage separately. This shared assignment prevents cross-stage leakage,
which is important because the assay stages overlap substantially: $997$ compounds appear at both
SP and DR, $159$ at both DR and MSP, and $97$ at all three stages. If the stages were split
independently, the same compound could appear in the test set at one stage and the training set at
another. This is particularly important for MSP-held: all $24$ MSP test compounds also have DR
measurements, and $15$ also have SP measurements. Without a shared split, a model trained jointly
across the cascade could therefore encounter every held-out MSP compound through earlier-stage
measurements during training.

Compounds were represented using radius-2 Morgan fingerprints, and chemically related compounds
were grouped into chemotype groups. Two compounds were assigned to the same connected chemotype
group if they shared a generic Bemis--Murcko framework \citep{bemis1996properties} or had a
Tanimoto similarity of at least $0.90$. Entire groups, rather than individual compounds, were then
assigned to the train, validation, or test sets. We preferentially selected the most chemically
isolated groups for the test set, using each compound's maximum fingerprint similarity to compounds
outside its own group to quantify isolation. Validation compounds were selected from the remaining
groups using the same criterion. This procedure produces deliberately challenging retrospective
evaluation sets while ensuring that closely related compounds are not divided between training and
evaluation.

Because the same global assignment is applied at every assay stage, the resulting dataset sizes
reflect both the different numbers of measurements available at each stage and their overlap. The
final splits contain $21{,}228$ training, $2{,}654$ validation, and $2{,}656$ test compounds at SP;
$1{,}273$, $163$, and $180$ compounds at DR; and $114$, $25$, and $24$ compounds at MSP,
respectively. For each test compound we compute its Tanimoto similarity to the most similar training compound.
Across test compounds, the median of this value is $0.643$, $0.634$, and $0.690$ at SP, DR, and
MSP respectively, and its largest value is $0.897$, $0.828$, and $0.828$. 

The prospective MSP evaluation set (MSP-prospective) was not used in constructing these splits. It
consists of the $84$ compounds selected using the single-task models, subsequently synthesized and
characterized experimentally, and therefore provides a separate evaluation from the retrospective
MSP-held set. Relative to the retrospective training set, MSP-prospective has a median maximum Tanimoto
similarity of $0.659$ and a maximum of $0.873$, placing it in a similar chemical-similarity regime
to the deliberately held-out retrospective sets.

\subsection{Stage-wise loss balancing}
\label{si:balancing}

The multi-task model combines the SP, DR, and MSP objectives using stage-wise loss balancing.
At optimization step $t$, an exponential moving average of the raw loss for each stage is updated as
\[
\bar{\mathcal{L}}_s^{(t)}
=
(1-\beta)\bar{\mathcal{L}}_s^{(t-1)}
+
\beta \mathcal{L}_s^{(t)},
\qquad \beta = 0.05,
\]
where $\mathcal{L}_s^{(t)}$ is the raw loss for stage $s$ at the current step and the moving
average is initialized from the first observed loss for each stage. The corresponding normalization
factor is
\[
a_s^{(t)}
=
\operatorname{clip}
\left(
\frac{1}{\left|\bar{\mathcal{L}}_s^{(t)}\right|},
\,0.1,\,10
\right),
\]
and the total multi-task objective is
\[
\mathcal{L}^{(t)}
=
\sum_{s \in \{\mathrm{SP},\,\mathrm{DR},\,\mathrm{MSP}\}}
w_s\,a_s^{(t)}\,\mathcal{L}_s^{(t)}.
\]
A loss is computed for each stage at every optimization step, so the moving averages are updated
continuously throughout training. The normalization factor $a_s^{(t)}$ is treated as a constant and
is not differentiated.

The SP, DR, and MSP objectives use different loss
functions and can therefore differ in scale. Magnitude normalization
and stage weights control their relative contributions to the
shared objective. At the start of training, their magnitudes differ
by more than an order of magnitude, reflecting the different loss functions used at each assay
stage rather than their relative importance. A separate diagnostic using BCE/Huber losses
illustrates how SP can dominate the scalar training loss
without balancing (Supplementary Figure~\ref{fig:reconstructed-stage-losses}).

The stage weights $w_s$ are fixed within a training run and selected by hyperparameter optimization
(Supplementary Section~\ref{si:hpo}). They are applied after magnitude normalization and control
the relative emphasis placed on each stage once the raw losses have been placed on a comparable
scale. The selected multi-task model uses $w_{\mathrm{SP}}=3.0$, $w_{\mathrm{DR}}=1.5$, and
$w_{\mathrm{MSP}}=2.0$.

Because the Gaussian negative log-likelihoods used at DR and MSP are not bounded below by zero,
their moving averages can approach or cross zero during training. We therefore normalize by the
absolute value of the moving average and clip the reciprocal normalization factor to $[0.1,10]$.
This prevents values near zero from producing unstable updates.

In the ``MT$-$loss balancing'' configuration,
moving-average normalization is disabled and all stage weights
are set to $w_s=1$. Its objective therefore reduces to the
unweighted sum of the three raw stage losses,
\[
\mathcal{L}_{\mathrm{unbalanced}}
=
\mathcal{L}_{\mathrm{SP}}
+
\mathcal{L}_{\mathrm{DR}}
+
\mathcal{L}_{\mathrm{MSP}}.
\]
For this configuration, checkpoint selection
also used the unnormalized mean validation score rather than
the chance-normalized criterion used for the multi-task model. The configurations were tuned independently and
also differ in their selected architecture and training
hyperparameters (Supplementary Tables~\ref{tab:hpo-selected-arch}
and~\ref{tab:hpo-selected-loss}).

\subsection{Exploratory training-loss comparison}

\begin{figure}[htbp]
  \centering
  \includegraphics[width=\linewidth]
    {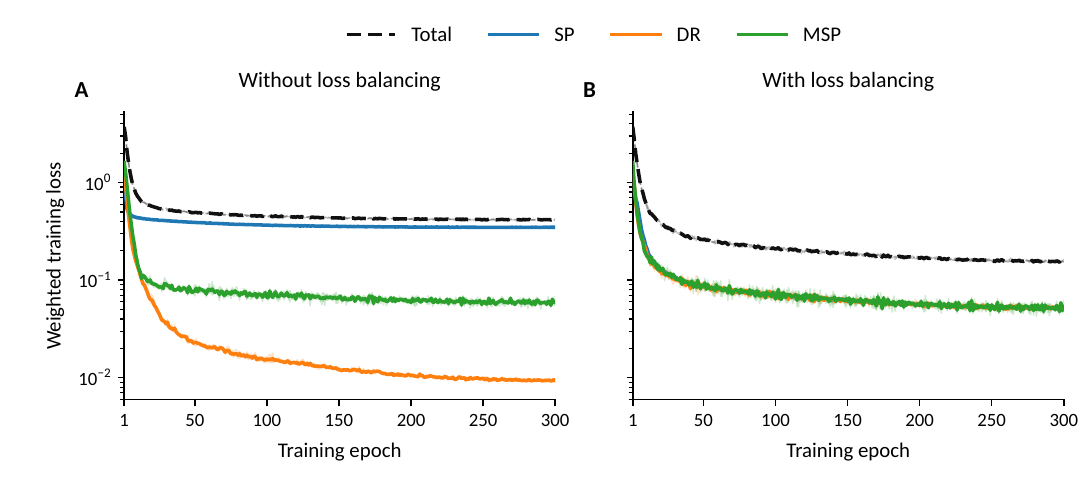}
  \caption{
  \textbf{Stage contributions to training loss in an exploratory
  multi-task model.}
  Total training loss (black dashed) and its SP, DR, and MSP
  contributions without (A) and with (B) epoch-wise geometric-mean
  loss balancing. Thin curves show three paired initializations
  and thick curves their means over all 300 epochs. Loss
  contributions are recorded during training and sum to the total;
  the vertical axis is logarithmic. The model uses soft-label
  binary cross-entropy for SP and standardized Huber losses for
  DR and MSP, distinct from the likelihood-based models in the
  main results.}
  \label{fig:reconstructed-stage-losses}
\end{figure}

To examine how loss scaling affects the shared objective,
we trained an exploratory multi-task model using soft-label binary
cross-entropy for SP and Huber loss on training-standardized DR and MSP
targets. Each stage loss was the sum of its per-target mean losses.
We compared unit weights with geometric-mean inverse-loss weights,
$w_s=G/L_s$, where $L_s$ is the mean stage loss from the preceding epoch
and $G$ is the geometric mean of the three stage losses. Weights were
initialized to one. Both configurations used the same training partition
and three paired initializations, with otherwise identical settings over
300 epochs. SP accounted for 82.2\% of the total training loss on average
across updates in epochs 101--200, compared with 33.8\% after balancing
(Figure~\ref{fig:reconstructed-stage-losses}). Although SP contributed
most of the scalar loss, its share of the summed shared-backbone gradient
norms was 14.5\% without balancing. These diagnostic runs use a different
loss formulation from the likelihood-based multi-task configurations
evaluated in Table~\ref{tab:generations}. Those configurations were tuned
independently and differ in their molecular representations, network
architectures, and checkpoint-selection criteria. Their performance
differences therefore cannot be attributed to loss balancing alone.

\subsection{Hyperparameter optimization}
\label{si:hpo}

Each model reported in Table~\ref{tab:generations} was tuned independently using the same budget
of 40 Optuna trials \citep{akiba2019optuna}. Searches used a tree-structured Parzen estimator with
a fixed sampler seed for reproducibility.

\paragraph{Neural-network search}
For hyperparameter optimization, each trial was trained for up to 12 epochs. Unpromising trials
were terminated using Optuna's median pruner with 3 start-up trials and 2 warm-up steps. Between
26 and 35 of the 40 trials were pruned for each model. The search space is given in
Table~\ref{tab:hpo-space}. All neural models use Morgan fingerprints, with fingerprint radius and
dimensionality optimized jointly with the network hyperparameters. Each single-task network and each multi-task
configuration received an independent hyperparameter search. The configuration with the highest validation
score among completed trials was selected for final training.

\begin{table}[htb]
\centering
\small
\begin{tabular}{lll}
\toprule
Hyperparameter & Range & Scale \\
\midrule
Fingerprint radius
    & $\{2, 3\}$ & categorical \\
Fingerprint length
    & $\{2048, 4096\}$ & categorical \\
Backbone width
    & $\{512, 1024, 2048\}$ & categorical \\
Residual blocks
    & $\{1, 2, 3\}$ & categorical \\
Dropout
    & $[0.05,\,0.35]$ & uniform \\
Head width
    & $\{0, 128, 256\}$ & categorical \\
Learning rate
    & $[5\times10^{-5},\,3\times10^{-3}]$ & log-uniform \\
Weight decay
    & $[10^{-6},\,10^{-3}]$ & log-uniform \\
$\lambda_{\sigma}$ (uncertainty loss)
    & $[0,\,1]$ & step $0.1$ \\
$\lambda_{\mathrm{act}}$ (activity loss)
    & $[0,\,1]$ & step $0.1$ \\
Stage weights $w_{\mathrm{SP}},w_{\mathrm{DR}},w_{\mathrm{MSP}}$
    & $[0.5,\,4.0]$ & step $0.5$ \\
\bottomrule
\end{tabular}
\caption{Hyperparameter search space for the neural models. Stage weights are optimized only for
the multi-task model; the MT $-$ loss balancing configuration uses equal stage weights by construction.
The linear warm-up period was fixed at 8 epochs for all neural models.}
\label{tab:hpo-space}
\end{table}

\paragraph{Model selection.}
Because the validation objectives differ across assay stages, configurations were ranked using a
chance-normalized mean validation score. Each stage's validation score $v$ was rescaled as
$(v-c)/(1-c)$, with $c=0.5$ for SP AUROC and $c=0$ for the DR and MSP correlations. For models
trained on multiple stages, the rescaled scores were averaged across stages. This places chance
performance at zero for each metric before averaging. The same criterion was used for checkpoint
selection during final training. The configuration without loss balancing
instead used the unnormalized mean validation score for
model selection.

\paragraph{Final training.}
The selected configuration for each neural model was retrained from scratch using five independent
random initializations. Models were trained for up to 300 epochs, with validation every 4 epochs
and early stopping after 20 validation checks without improvement, subject to a minimum of
30 epochs. In practice, training terminated between epochs 88 and 252. For each run, the checkpoint with the highest
validation score under the selection criterion used for that
configuration was restored for evaluation.
Table~\ref{tab:generations} reports the mean and standard deviation across the five runs.

\begin{table}[htb]
\centering
\small
\setlength{\tabcolsep}{6pt}
\begin{tabular}{lrrrrrr}
\toprule
Model & Radius & Bits & Width & Blocks & Head & Dropout \\
\midrule
Single-task (SP)     & 3 & 4096 & 512  & 3 & 256 & 0.086 \\
Single-task (DR)     & 2 & 4096 & 1024 & 3 & 256 & 0.099 \\
Single-task (MSP)    & 3 & 4096 & 512  & 3 & 128 & 0.217 \\
MT $-$loss balancing & 2 & 2048 & 1024 & 1 & 256 & 0.196 \\
Multi-task (MT)      & 3 & 4096 & 2048 & 2 & 256 & 0.125 \\
\bottomrule
\end{tabular}
\caption{Selected molecular-representation and architecture hyperparameters for the neural models.
``Head'' denotes the width of the stage-specific prediction head.}
\label{tab:hpo-selected-arch}
\end{table}

\begin{table}[htb]
\centering
\small
\setlength{\tabcolsep}{5pt}
\begin{tabular}{lcccc}
\toprule
Model & Learning rate / WD &
$\lambda_{\sigma}$ & $\lambda_{\mathrm{act}}$ & Stage weights \\
\midrule
Single-task (SP)
& $8.5\times10^{-4}$ / $1.1\times10^{-6}$
& 0.6 & 0.7 & --- \\
Single-task (DR)
& $1.4\times10^{-3}$ / $5.1\times10^{-5}$
& 0.4 & 0.0 & --- \\
Single-task (MSP)
& $2.0\times10^{-3}$ / $4.2\times10^{-4}$
& 0.5 & 0.1 & --- \\
MT $-$loss balancing
& $7.6\times10^{-4}$ / $3.6\times10^{-5}$
& 0.5 & 0.2 & $1.0/1.0/1.0$ \\
Multi-task (MT)
& $2.6\times10^{-4}$ / $3.5\times10^{-4}$
& 0.5 & 0.5 & $3.0/1.5/2.0$ \\
\bottomrule
\end{tabular}
\caption{Selected optimization and loss hyperparameters for the neural models. ``WD'' denotes
weight decay. Stage weights are reported as
$w_{\mathrm{SP}}/w_{\mathrm{DR}}/w_{\mathrm{MSP}}$; MT $-$loss balancing uses equal stage weights
by construction.}
\label{tab:hpo-selected-loss}
\end{table}

\paragraph{Linear baselines}
The linear baselines were optimized using the same 40-trial budget. Morgan fingerprint radius and
dimensionality were searched over the same ranges used for the neural models, and the ridge
regularization strength was searched log-uniformly over $10^{-2}$--$10^{5}$. For Linear (all
stages), stage weighting was included as an additional hyperparameter. Selection used
macro-averaged validation Spearman correlation across the prediction targets available to each
model. Linear selected radius 3 with 2048 bits and $\alpha=1.9\times10^{3}$; Linear (all stages)
selected radius 2 with 4096 bits and $\alpha=6.7\times10^{4}$, in both cases without stage
weighting. Because the linear estimators are deterministic given their inputs and hyperparameters,
they are reported as single-fit point estimates rather than means across random initializations.

\subsection{Paired crude-DR and purified-MSP selectivity}
\label{si:paired-selectivity}

\begin{figure}[htbp]
  \centering
  \setlength{\fboxsep}{2pt}
  {\includegraphics[width=0.97\linewidth]{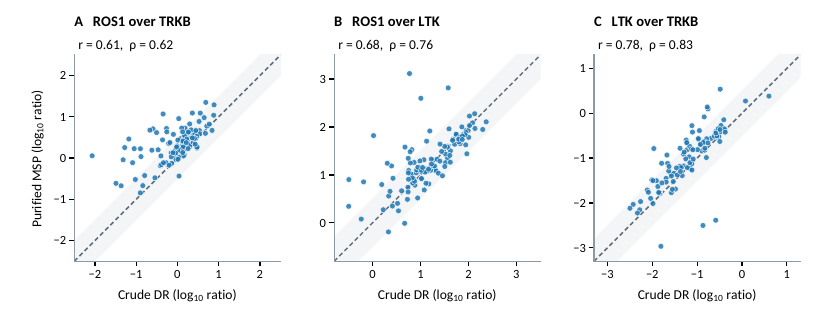}}
  \caption{\textbf{Paired crude and purified selectivity.}
  Compound-matched DR and MSP selectivity for \textbf{(A)} ROS1/TRKB,
  \textbf{(B)} ROS1/LTK, and \textbf{(C)} LTK/TRKB ($n=121$, 114, and 99).
  Points are pooled compound-level posterior summaries from all retrospective splits;
  both targets must be resolved at both stages. Dashed lines indicate equality and shading a
  descriptive threefold band. Pearson $r$ and Spearman $\rho$ are shown.}
  \label{fig:dr-msp-selectivity}
\end{figure}

We matched the submitted DR and historical MSP datasets by canonical molecular structure,
yielding 159 one-to-one compound pairs. Each target pair was evaluated only when all four
stage/target potency summaries were finite and resolved ($p_{\mathrm{inact}}<0.30$), using the
same cutoff as the main evaluation. For targets $A$ and $B$, selectivity for $A$ over $B$ is
$\log_{10}[\mathrm{IC}_{50}(B)/\mathrm{IC}_{50}(A)]$; positive values favor $A$.
The eligible counts are 121 for ROS1/TRKB, 114 for ROS1/LTK, and 99 for LTK/TRKB.
Their Pearson correlations are 0.61, 0.68, and 0.78; Spearman correlations are 0.62, 0.76, and
0.83, respectively (Figure~\ref{fig:dr-msp-selectivity}).

These associations support approximate retention of selectivity in the selected, resolved
subset. They do not establish exact equality: median MSP-minus-DR shifts in log selectivity
are $+0.22$, $+0.002$, and $+0.18$, respectively. Posterior summaries pool measurements
within each stage, and MSP entails resynthesis, purification, and quantification rather than
purification of the same assayed aliquot. Differences may therefore reflect multiple experimental
effects. The analysis neither excludes target-specific crude-mixture interference nor establishes
reliability for compounds not advanced to MSP.

\begin{figure}[htbp]
  \centering
  \setlength{\fboxsep}{2pt}
  \includegraphics[width=0.60\linewidth]{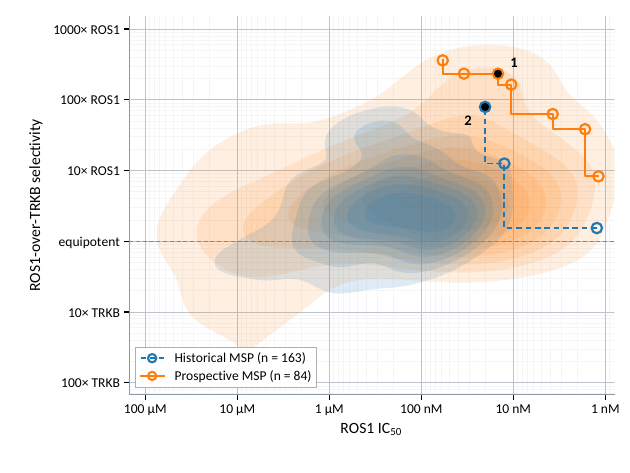}
\caption{\textbf{MSP-only counterpart to Figure~\ref{fig:best}B.}
Blue and orange contours show the historical ($n=163$) and prospective ($n=84$)
purified MSP cohorts, respectively. Open circles and connecting steps mark the
Pareto frontiers, which remain unchanged after excluding crude DR measurements.}
  \label{fig:pareto-msp-only}
\end{figure}

\subsection{MSP-only potency--selectivity landscape}
\label{si:purified-frontier}

Supplementary Figure~\ref{fig:pareto-msp-only} shows the MSP-only counterpart to
Figure~\ref{fig:best}B. Both Pareto frontiers remain unchanged because all three
historical frontier compounds were already measured by MSP; excluding DR changes
only the historical density.

The two cohorts differ in how their compounds were selected. The 163 historical
MSP compounds were chosen for purification after experimental SP/DR screening,
so this group was already enriched for activity. The 84 prospective compounds
were selected using model predictions and tested directly by MSP, without
experimental SP/DR screening. Their enrichment for activity was therefore
achieved without this experimental prefiltering. The historical MSP cohort
provides a useful reference for the compounds discovered, but is not a matched
control for evaluating model-based selection.

\section{Limitations}
\label{si:limitations}

This study focuses on a single 2,4-disubstituted pyrimidine
core within a kinase-focused discovery program. The extent to which these
findings generalize to other chemical scaffolds and target classes therefore
remains to be established. In addition, the purified evaluation sets are
small (24 retrospective and 84 prospective compounds before potency
filtering), limiting the precision of performance estimates and comparisons
between models. Further evaluation using larger purified datasets across
a broader range of chemistries and targets will be needed to establish
the wider applicability of the approach.

\appendix

\end{document}